\documentclass[]{interact}
\usepackage{color}
\usepackage{epstopdf}
\usepackage[caption=false]{subfig}

\usepackage[numbers,sort&compress,merge]{natbib}
\bibpunct[, ]{[}{]}{,}{n}{,}{,}

\theoremstyle{plain}

\theoremstyle{definition}

\theoremstyle{remark}

\begin{document}


\title{Kinetic theory of diffusion in a channel of varying cross section }

\author{
\name{J. Javier Brey\textsuperscript{a,b}\thanks{CONTACT J. Javier Brey. Email: brey@us.es}, M. I. Garc\'{\i}a de Soria\textsuperscript{a}, and P. Maynar\textsuperscript{a}}
\affil{\textsuperscript{a} F\'{\i}sica Te\'{o}rica, Universidad de Sevilla, Apartado de Correos 1065, E-41080, Sevilla, Spain; \textsuperscript{b} Institute for Theoretical and Computational Physics, Facultad de Ciencias, Universidad de Granada, E-18071, Granada, Spain}
}

\maketitle

\begin{abstract}
Self-diffusion along the longitudinal coordinate in a channel of varying cross section is considered. The starting point is the two-dimensional  Enskog-Boltzmann-Lorentz kinetic equation with appropriated boundary conditions. It is integrated over the transversal coordinate to get an approximated one-dimensional kinetic equation, keeping the relevant properties of the original  one.  Then, a macroscopic equation for the time  evolution of the longitudinal density is derived, by means of a modified Chapman-Enskog expansion method, that takes into account the inhomogeneity of the equilibrium longitudinal density. This transport equation  has the form of the phenomenological Ficks-Jacobs equation, but with an effective diffusion coefficient that contains corrections associated to the variation of the slope of the equilibrium longitudinal density profile.

\end{abstract}


\section{Introduction}
\label{s1}
There are many processes in nature  that can be accurately understood in terms  of a simplified one-dimensional description. This happens, in particular, when the motion of the particles that compose  the system is strongly restricted in every dimension except one, and the interest focusses on  transport processes along it. This paper deals with the derivation of macroscopic laws describing the one-dimensional dynamics in these systems, starting from a microscopic description at the level of kinetic theory. The simplest transport process of self-diffusion will be considered \cite{Do75}.

Diffusion in systems with geometrical constraints is an interesting problem often encountered in biological and soft matter systems.  Consider, for instance, a two-dimensional channel of varying cross section and that the relevant process is the diffusion of some solute or labeled particles along the tube. To be more precise, suppose a two-dimensional (2D) system confined by two boundaries located at $y=h(x)$ and  $y= g(x)$, respectively, being $h(x)>g(x)$, and let us denote by $n_{l}(x,y,t)$ the number density of labeled particles. Of course, the $x$ axis is taken along the open direction of the channel. Our aim is to get a closed  equation describing the time evolution of the projected one dimensional density of labeled particles along the $x$ axis, $n_{lx}(x,t)$, defined by
\begin{equation}
\label{1.1}
n_{lx}(x,t) \equiv \int_{g(x)}^{h(x)} dy\, n_{l}(x,y,t).
\end{equation}
A heuristic  derivation of an effective one-dimensional diffusion equation for this problem was given by Jacobs \cite{Ja67}, who attributed the idea to Fick \cite{Fi55}.  For this reason, the equation is usually referred to as the Fick-Jacobs (FJ) equation \cite{Zw92}, and it has the form
\begin{equation}
\label{1.2}
\frac{\partial}{\partial t} n_{l,x}(x,t)=D_{0} \frac{\partial}{\partial x} \left[ A(x) \frac{\partial}{\partial x} \frac{n_{lx}(x,t)}{A(x)} \right],
\end{equation}
where  $A(x) \equiv h(x)-g(x)$ is the width of the channel and $D_{0}$ the diffusion coefficient. Notice that if there were no change in the section of the channel, the above equation reduces to a simple diffusion equation, and that the correction due to a varying shape is of first order in $\partial A(x)/\partial x$.   To avoid misunderstandings, it must be stressed that the FJ equation focusses in describing the effects of the variation of the width of the channel on transport. In other words, it is assumed that the system is such that in the limit of a constant section, the equation correctly describes transport along the $x$ axis.  

A more fundamental derivation of the FJ equation was carried out by  Zwanzig \cite{Zw92}. His starting point was the two-dimensional Smoluchowski equation  for diffusion through a general potential.  In the derivation, Zwanzig  uses a local equilibrium assumption, namely that the relaxation in the transverse direction is infinitely fast, so that the transversal density profile  for a given value of $x$ is flat. He also analyzes the effect of the deviations from local equilibrium for systems with cylindrical symmetry, expressing his results in the form on an effective longitudinal diffusion coefficient, $D_{Z}(x)$, so that the modified FJ equation proposed in ref. \cite{Zw92} reads
\begin{equation}
\label{1.3}
\frac{\partial}{\partial t}n_{l,x}(x,t) = \frac{\partial}{\partial x} \left[ D_{Z}(x) A(x) \frac{\partial}{\partial x} \frac{n_{lx}(x,t)}{A(x)} \right].
\end{equation}
The coefficient $D_{Z}(x)$ is estimated  from truncated expansions and expressed as a function of $ \left ( \partial A (x) / \partial x \right)^{2}$. To zeroth order in this quantity it reduces to $D_{0}$.

Diffusion in a quasi-one-dimensional channel is also considered in \cite{KyP05,KyP06}. In these papers, the authors start from the anisotropic two-dimensional diffusion equation, with different diffusion coefficients in the longitudinal ($D_{x}$) and transverse ($D_{y}$) directions. The equation is supplemented with reflecting  boundaries conditions at the two confining walls. Then, a systematic expansion in the ratio $D_{x}/D_{y}$ is considered, the lowest order approximation corresponding to the FJ equation. Higher order corrections can be expressed in terms of the expansion of an effective diffusion coefficient $D_{KP}(x)$. 
The expansion technique used is approximative: it is assumed that the transverse relaxation processes are fast enough as to track the longitudinal ones and to  generate a  ``local steady state''  \cite{KyP05,KyP06}.

Using mesoscopic nonequilibrium thermodynamic methods \cite{GyM84}, Reguera and Rub\'{\i} \cite{RyR01} derived a kinetic equation describing diffusion when the statics of the system is characterized  by a thermodynamic potential landscape. The equation is applied to the case of diffusion in a channel of varying cross section and the FJ equation is found. Moreover, a scaling law for the effective diffusion coefficient is formulated by the authors on the basis of rather heuristic arguments, resulting in an expression $D_{RR}(x)$, which is also a function of $(\partial A(x) / \partial x)^{2}$. Let us notice that the three expressions, $D_{Z}(x)$, $D_{KP}(x)$, and $D_{ZZ}(x),$ are different  beyond  the first order correction.

Our purpose in this paper is to consider again the same problem of diffusion  in a channel of varying cross section, but starting from a description of the system at the particle level and using kinetic theory methods. In addition to the interest of, eventually,  providing a more microscopic basis to the FJ equation, there are several reasons that make this study relevant. Firstly, in the FJ equation the flux of particles appears as proportional to the gradient of the density scaled with the channel width.  This is not the usual form obtained from kinetic theory by using any of the standard methods employed to derive transport equations, e.g. the Chapman-Enskog procedure, in which the hydrodynamic fluxes are developed in an uniformity parameter associated to the order in the gradients of the hydrodynamic fields \cite{RydL77}. Secondly, the derivations mentioned above start from a two-dimensional transport equation which is consistent up to second order in the density gradient. When this equation is integrated over the transversal coordinate, it is not clear that the result be consistent up to a given order in the gradient of $n_{lx}(x,t)/A(x)$. In this context, it is important to realize that $A(x)$ is proportional to the equilibrium value of $n_{lx}$, as a consequence of the homogeneity of the equilibrium state. The analysis to be reported here  indicates the presence of an effective diffusion coefficient depending on $\partial^{2} A(x)/\partial x^{2}$, that is conceptually different from the dependence found in previous studies \cite{Zw92,KyP05,KyP06,RyR01}. A third reason for the present work is that the use of two-dimensional macroscopic equations implies assuming that hydrodynamics also holds in the transversal direction, that requires, in particular, that the separation of the two boundaries be much larger than the mean-free-path of the particles for all $x$. The analysis performed in this paper shows that this is not the case; to get a closed equation for the transport of particles along the longitudinal direction, hydrodynamics is only needed along that direction. Quite peculiarly, the original heuristic derivation given by Jacobs \cite{Ja67}  is based on the one-dimensional Fick law, and not in a reduction of the number of variables in an equation for a larger dimension. Let us also mention that the seminal idea of Jacobs has been also employed to study diffusion of binary mixtures under nano-confinement \cite{MMyP15}.

The remaining of this paper is distributed in the following way. In Sec. \ref{s2}, the one-dimensional kinetic equation for the one-particle distribution function of labeled particles along the $x$ axis is derived from the two-dimensional Enskog-Boltzmann-Lorentz equation for the self-diffusion of labeled particles in an equilibrium system confined by elastic hard boundaries. To get a closed one-dimensional kinetic equation, it is necessary to introduce some approximation about the behavior of the two-dimensional distribution function at the boundaries. A simple choice compatible with the boundary conditions and keeping the main properties of the exact kinetic equation is made. The physical meaning of the approximation is analyzed and related with the idea of entropic barrier introduced by Zwanzig \cite{Zw92,ZyZ91}. Next, in order to obtain a normal solution of  the kinetic equation leading to a hydrodynamic description, a modified Chapman-Enskog expansion is formulated in Sec. \ref{s3}. It is based on an expansion in a uniformity parameter associated to the gradients of the actual density field relative to the equilibrium density field \cite{BGyM20}. The underlying idea in this expansion is that, since the equilibrium density along the longitudinal direction  is not homogeneous, the existence of a density gradient is not enough to imply the presence of a macroscopic flow of particles in the system. 

The modified Chapman-Enskog expansion is implemented in Sec. \ref{s4} to first order in the uniformity parameter. The resulting transport equation for the longitudinal density of labeled particles can be written as a  FJ equation, with an effective diffusion coefficient that contains a term proportional to $\partial ^{2} A(x)/\partial x^{2}$. This differs from the results obtained in  previous analysis of the problem, in which the effect of transversal inhomogeneities is investigated in the context of two-dimensional macroscopic transport equations. The paper ends in Sec. \ref{s5} with a short summary of the main conclusions. 
Appendices \ref{ap1} and \ref{ap2} contain details of calculations leading to results mentioned in the main body of the paper.

It is a pleasure to dedicate this work to our colleague and friend Luis Rull, with whom we have shared many discussions of physics (and also almost everything else).

\section{The one-dimensional kinetic equation}
\label{s2}
Consider a gas composed of hard disks  of mass $m$ and diameter $\sigma$.  The system is at equilibrium at temperature $T$, and the number of particles density is $n$.  Then, the one-particle distribution function, $f_{eq}({\bm v})$, has the form
\begin{equation}
\label{2.1}
f_{eq}({\bm v}) = n \varphi ({\bm v}),
\end{equation}
where $\varphi({\bm v})$ is the Maxwellian velocity distribution,
\begin{equation}
\label{2.2}
\varphi({\bm v}) = \frac{m}{2 \pi k_{B}T}  e^{-\frac{mv^{2}}{2 k_{B}T}},
\end{equation}
$k_{B}$ being the Boltzmann constant. Suppose now that, although all particles are mechanically equivalent,  some of them are labeled or have a different ``color'' than the others. The one-particle distribution function of colored particles will be denoted by $f_{l} ({\bm r},{\bm v},t)$. Of course, if the system as a whole was initially at equilibrium, it will remain at equilibrium, independently of the initial distribution of labeled particles. On the other hand, if the distribution function of labeled particles is not the equilibrium one, it will evolve in time until reaching the same expression as given in Eq.\ (\ref{2.1}), but with the average density of labeled particles $n_{l}$, instead of the total density $n$.

The Enskog theory provides a successful empiric kinetic equation to study gases of hard particles at low and moderate densities \cite{RydL77,McL89}. Applied to the present case, it reads
\begin{equation}
\label{2.3}
\left( \frac{\partial}{\partial t}
 +{\bm v} \cdot {\bm \nabla} \right) f_{l}({\bm r}, {\bm v}, t) =  g_{e}(n) \Lambda_{BL} [{\bm v}|f_{eq}] f_{l} ({\bm r}, {\bm v},t).
 \end{equation}
Here $g_{e}(n)$ is the equilibrium pair correlation of two particles of the gas at contact and $\Lambda_{BL}$ is the Boltzmann-Lorentz (BL) operator defined by \cite{RydL77}
\begin{equation}
\label{2.4}
\Lambda_{BL}[{\bm v}|f_{eq}] \psi ({\bm v}) \equiv \sigma \int d{\bm v}_{1} \int  d \widehat{\bm \sigma}\,   |{\bm g} \cdot \widehat{\bm \sigma} | \Theta ( {\bm g} \cdot \widehat{\bm \sigma} ) \left[ f_{eq} ({\bm v}^{\prime}_{1}) \psi ({\bm v}^{\prime})- f_{eq} ({\bm v}_{1}) \psi ({\bm v}) \right],
\end{equation}
for arbitrary $\psi({\bm v})$. 
In this expression, ${\bm g} \equiv {\bm v}_{1} - {\bm v}$ is the relative velocity of the two collisional particles prior to the collision, $d \widehat{\bm \sigma}$ is the solid angle element around the unit vector $\widehat{\bm \sigma}$ joining the centers of the two particles at contact, and $\Theta$ is the Heaviside step function. Finally, ${\bm v}^{\prime}$ and ${\bm v}^{\prime}_{1}$ are the postcollisional velocities given by
\begin{eqnarray}
\label{2.5}
{\bm v}^{\prime} & =  &{\bm v} + {\bm g} \cdot \widehat{\bm \sigma}  \widehat{\bm \sigma}, \nonumber \\
{\bm v}^{\prime}_{1} & =  &{\bm v}_{1} - {\bm g} \cdot \widehat{\bm \sigma}  \widehat{\bm \sigma}.
\end{eqnarray}
Upon writing  Eq.\ (\ref{2.4}), it has been used that labeled particles collide with both labeled and unlabeled particles. 
In the present context, there is no difference between the original equation formulated by Enskog \cite{CyC70} and the revised theory introduced by van Beijeren and Ernst \cite{vByE73}. In both cases, the local density at which the equilibrium pair correlation function must be evaluated is the uniform total one $n$, since it determines the collision frequency of any particle. 

For an arbitrary pair of velocity functions, $\psi({\bm v})$ and $\chi({\bm v})$, it is
\begin{equation}
\label{2.6}
\int d{\bm v}\,  \chi({\bm v}) \Lambda_{BL}[{\bm v}|f_{eq}] \psi({\bm v})= \sigma \int d{\bm v} \int d{\bm v}_{1} \int d \widehat{\bm \sigma}\,  |{\bm g} \cdot \widehat{\bm \sigma} | \Theta ( {\bm g} \cdot \widehat{\bm \sigma} ) f_{eq}({\bm v}_{1})  \psi({\bm v}) [ \chi({\bm v}^{\prime})- \chi ({\bm v}) ].
\end{equation}

\begin{figure}
\centering
\includegraphics[scale=0.6,angle=0]{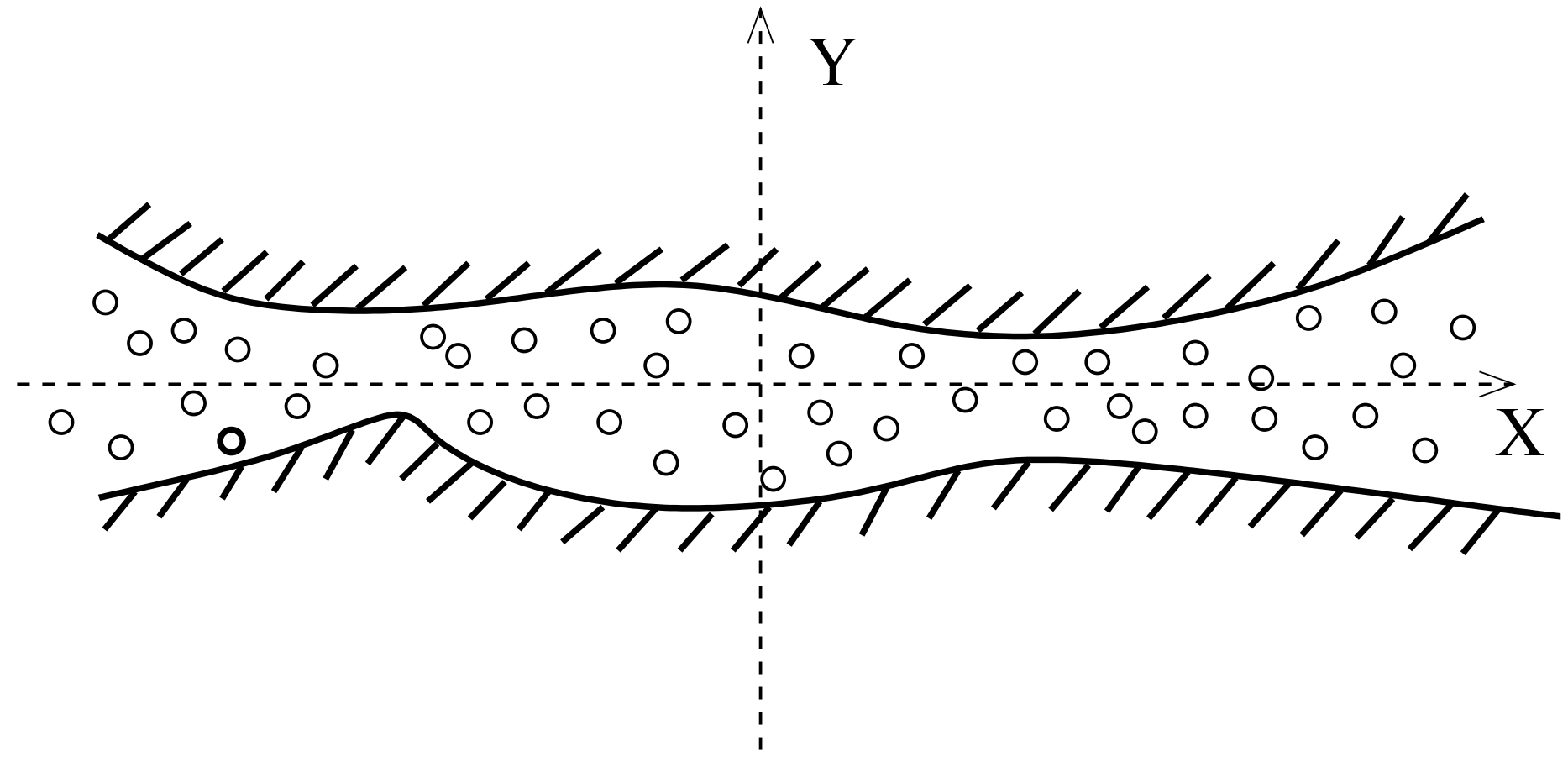}
\caption{ Sketch of a channel with varying cross section. The origin of the vertical axis is irrelevant for the results derived in this paper.}
\label{fig1}
\end{figure}

The aim of this paper is to study the diffusion of labeled particles in a channel of varying cross section. The system under consideration is confined by elastic hard walls except in one direction, taken as the $x$-axis. More specifically, consider that    the two hard boundaries  are defined by the curves $y=h(x)$ and $y=g(x)$, $h(x)>g(x)$, for all $x$. It will be shown that the results do not depend on the origin taken for the perpendicular $y$-axis. A sketch of the system is given in Fig. \ref{fig1}. A marginal one-particle distribution function of labeled particles along the $x$  direction, $f_{lx}(x,{\bm v},t)$, is defined by
\begin{equation}
\label{2.8}
f_{lx}(x,{\bm v},t) \equiv \int_{g(x)} ^{h(x)}dy\, f_{l}(x,y,{\bm v},t).
\end{equation}

Then, on integrating the two-dimensional equation (\ref{2.3}), it is obtained
\begin{eqnarray}
\label{2.9}
&&\left( \frac{\partial}{\partial t}
 +v_{x} \frac{\partial}{\partial x} \right) f_{lx}(x, {\bm v}, t) - v_{x} \frac{\partial h(x)}{\partial x} f_{l}(x,h(x),{\bm v},t) + v_{x} \frac{\partial g(x)}{\partial x} f_{l}(x,g(x),{\bm v},t) \nonumber \\
 &&+ v_{y} \left[ f_{l}(x,h(x),{\bm v},t) - f_{l}(x,g(x),{\bm v},t) \right]  = g_{e}(n) \Lambda_{BL} [{\bm v}|f_{eq}] f_{lx} (x, {\bm v},t).
 \end{eqnarray}
 The average number density of labeled particles per unit of length along the $x$-axis is
 \begin{equation}
 \label{2.10}
 n_{lx}(x,t) \equiv \int d{\bm v}\, f_{lx}(x,{\bm v},t).
 \end{equation}
 A conservation law follows by velocity integration of Eq.\  (\ref{2.9}),
 \begin{equation}
 \label{2.11}
 \frac{\partial}{\partial t}\, n_{lx}(x,t)+ \frac{\partial}{\partial x}\, J_{l}(x,t)=0,
 \end{equation}
 where $J_{l}(x,t)$ is the flux of labeled particles along the $x$ direction,
 \begin{equation}
 \label{2.12}
 J_{l}(x,t) \equiv  \int d{\bm v}\, v_{x} f_{lx}(x,{\bm v},t).
 \end{equation}
 Notice that $J_{l}(x,t)$ is not actually a flux, but a ``caudal'', in the sense that it is the number of labeled particles that cross the section located at $x$ by unit of time.  In the derivation of Eq.\ (\ref{2.11}), use has been made of the kinetic boundary conditions at the elastic hard walls confining the system. Details of the calculations are provided en Appendix \ref{ap1}. The kinetic equation (\ref{2.3}) with the assumed boundary conditions has the equilibrium solution
 \begin{equation}
 \label{2.13}
 f_{l,eq}({\bm r} ,{\bm v} ) = n_{l}  \varphi({\bm v}),
 \end{equation}
 for $g(x) <y<h(x)$, and zero otherwise. The corresponding equilibrium form for the marginal distribution along the $x$ axis is obtained by means of Eq. (\ref{2.8}),
 \begin{equation}
 \label{2.14}
 f_{lx,eq}(x,{\bm v}) = A(x)  n_{l}  \varphi({\bm v}).
 \end{equation}
 It is trivial to verify that the above expression together with Eq.\ (\ref{2.13}) applied to the hard boundaries define a steady solution of Eq.\ (\ref{2.9}), as required for consistency. The equilibrium density of labeled particles along the $x$ direction is
 \begin{equation}
 \label{2.15}
 n_{lx,eq} = A(x) n_{l}\, .
 \end{equation}
 
 To convert Eq. (\ref{2.9}) into a closed equation for $f_{lx}(x,{\bm v},t)$, it is necessary to make some approximation on the boundary values  of the complete two-dimensional distribution, $f_{l}(x,h(x),{\bm v},t)$ and $f_{l} (x,g(x),{\bm v},t)$. They must be expressed in terms of the one-dimensional distribution. Any sensible simplified form of the equation must verify the following two conditions: (i) it must keep the form of the exact conservation law, Eq. (\ref{2.11}), including the expression of the flux of labeled particles in terms of the marginal distribution function, and (ii) it must admit the equilibrium distribution given by Eq. (\ref{2.14}). Here, the following approximation will be considered:
 \begin{equation}
 \label{2.16}
 f_{l} \left( x,h(x),{\bm v},t \right)   \approx  f_{l} \left( x,g(x),{\bm v},t \right) \approx  \frac{f_{lx,0}(x,{\bm v},t)}{A(x)}\, 
 \end{equation}
 where
 \begin{equation}
 \label{2.16a}
 f_{lx,0} (x,{\bm v},t) \equiv n_{lx}(x,t) \varphi ({\bm v}) 
 \end{equation}
 is the local equilibrium distribution of labeled particles along the $x$ axis. The above condition equates the densities at the points $x,h(x)$ and $x,g(x)$, plus set these equal to the average density on the line connecting them. This is expected to be a good approximation if the derivatives of $g$ and $h$ with respect to $x$ remain small in absolute value in units of the mean free path of the particles. Then, the kinetic equation (\ref{2.9}) becomes
 \begin{equation}
 \label{2.17}
 \left( \frac{\partial}{\partial t}
 + v_{x} \frac{\partial}{\partial  x} \right) f_{lx}(x, {\bm v}, t)  - v_{x} \frac{\partial \ln n_{lx,eq}(x)}{\partial x}\, f_{lx,0}(x,{\bm v},t) 
 =  g_{e}(n) \Lambda_{BL} [{\bm v}|f_{eq}] f_{lx} (x, {\bm v},t).
 \end{equation}
 This equation is easily seen to fulfill the two conditions enumerated above. It is worth to stress that the approximation made only refers to the value of the two-dimensional one-particle distribution function of labeled particles at the hard boundaries, i.e., at $y=g(x)$ and $y=h(x)$. Homogeneity along the transversal $y$ direction has not been assumed. Another relevant point is that the approximation formulated in Eq.\ (\ref{2.16}) verifies the exact kinetic boundary conditions of the two-dimensional Lorentz-Enskog equation as formulated in Eqs.\ (\ref{ap1.2}) and (\ref{ap1.4}). In any case, it is clear that Eq. (\ref{2.16}) is a lower order approximation, which can be improved on the basis of a more detailed analysis of the two-dimensional Enskog-Boltzmann-Lorentz equation.
 
 It is interesting to analyze the second term on the left-hand side of Eq. (\ref{2.17}), since it contains the effects of the varying cross section on the effective dynamics of the gas in the longitudinal direction. This term can be written in the equivalent form
 \begin{equation}
 \label{2.18}
 \frac{k_{B}T}{m} \left[ \frac{\partial}{\partial x} \ln   A(x)\right] \frac{\partial}{\partial v_{x}}  f_{lx,0} (x, {\bm v},t).
 \end{equation}
 This is formally similar to a term describing the effect of an  external field $F_{x}$ acting on the particles in the direction of the $x$ axis.  In general, the latter reads \cite{RydL77,McL89}
 \begin{equation}
 \label{2.19}
 \frac{F_{x}}{m} \frac{\partial}{\partial v_{x}} f_{lx}(x,{\bm v},t).
 \end{equation}
 To establish a correspondence between Eqs. (\ref{2.18}) and (\ref{2.19}),  the local equilibrium approximation must be made,
 \begin{equation}
 \label{2.20}
 f_{lx}(x,{\bm v},t)  \approx  f_{lx,0}(x,{\bm v},t)
 \end{equation}
 and the external force must be identified as deriving from the potential
 \begin{equation}
 \label{2.21}
 U(x) = -k_{B}T \ln \frac{A(x)}{l_{0}},
 \end{equation}
$l_{0}$ being some characteristic length, for instance the mean free path of the particles.  An alternative of the above expression for the potential associated to the force, also leading to the expression given in Eq. (\ref{2.18}) is
 \begin{equation}
 \label{2.22}
 \Phi(x) = - T S(T,x).
 \end{equation}
 Here
 \begin{equation}
 \label{2.23}
 S(T,x)= k_{B} \ln \Omega (T,x)
 \end{equation}
 and $\Omega(x)$ the phase space accesible to the original two-dimensional system per unit of length along the longitudinal axis, that is simply proportional to the width $A(x) = h(x)-g(x)$. Formally, $\Omega$ is assumed to be measured in some dimensionless units. The expression in Eq.\ (\ref{2.23}) agrees with the microcanonical definition of entropy. Then, $\Phi(x)$ can be interpreted as an entropy potential of geometrical origin, i.e. associated to the space available for diffusing particles. This idea of associating the effect of a varying cross section with an entropy barrier was introduced by Zwanzig \cite{Zw92}, who also emphasized that it does not means that the entropy is lower at the barrier, but the latter is associated solely with changes in entropy.

\section{Modified Chapman-Enskog expansion and the zeroth order distribution function}
\label{s3}
Once the model kinetic equation has been specified, our next goal is to derive a diffusion equation from it, i.e. a closed evolution equation for the one-dimensional number density of labeled particles, $n_{lx}(x,t)$.  This requires to express the flux $J_{l}(x,t) $ appearing  in Eq. (\ref{2.11}) as a functional of the density, i.e. as a function of $n_{lx}(x,t)$ and its spatial derivatives. To do so, a modification of the well-known Chapman-Enskog procedure \cite{McL89,CyC70} will be employed. The basic starting idea of the method is to look for a ``normal'' solution of the kinetic equation. By definition, a normal distribution function is such that all its space and time  dependence occurs through the macroscopic fields and their gradients. In the present case, the only macroscopic field is the number density of labeled particles,  so that a normal distribution has the form
\begin{equation}
\label{3.1}
f_{lx}(x,{\bm v},t)= f \left( {\bm v}, n_{lx}(x,t), \frac{\partial}{\partial x} n_{lx}(x,t), \frac{\partial^{2}}{\partial x^{2}} n_{lx}(x,t), \ldots \right).
\end{equation}
The normal distribution is therefore a functional of $n_{lx}(x,t)$. In the following, the above type of functional dependence will be denoted in the compact form
\begin{equation}
\label{3.2}
f_{lx}(x,{\bm v},t)=f_{lx} \left[ {\bm v}|n_{lx}(x,t) \right].
\end{equation}
To find the normal solution, the usual Chapman-Enskog algorithm  applied to a bulk system employs an expansion in the gradients of the hydrodynamic fields. The underlying assumption is that the deviations of these fields from their equilibrium 
values are smooth in the sense that their relative variation over a distance of the order of the mean free path is small. The expansion of the distribution function generates a similar expansion for the flux of the conserved quantities. Nevertheless, in the self-diffusion problem being analyzed here, 
the equilibrium distribution of labeled particles along the $x$ axis, $n_{lx,eq}$,  is inhomogeneous.  As a consequence, only the gradients associated to the deviations of the density from  its equilibrium value are expected to generate a flux of 
labeled particles, and the gradient expansion of $n_{lx,eq}(x)$ must be avoided. To formulate the perturbation scheme to be employed in the following, a dimensionless function $\nu(x,t)$ is defined by the relation \cite{BGyM20}
\begin{equation}
\label{3.3}
n_{lx}(x,t) \equiv n_{lx,eq}(x) \nu(x,t).
\end{equation}
At equilibrium, it is $\nu=1$. Next, a formal uniformity parameter $\epsilon$ is introduced. It will be associated to the gradient of $\nu(x,t)$, but not of $n_{lx,eq}(x)$. More specifically, the gradient of the density of labeled particles is
 decomposed in the form
\begin{equation}
\label{3.4}
\frac{\partial n_{lx}(x,t)}{\partial x} = \partial^{(0)}_{x} n_{lx}(x,t)+ \epsilon \partial_{x}^{(1)} n_{lx}(x,t),
\end{equation}
where the two operators on the right hand side are defined by 
\begin{equation}
\label{3.5}
\partial_{x}^{(0)} n_{lx} (x,t) \equiv  \nu (x,t) \frac{\partial}{\partial x}\, n_{lx,eq}(x)
\end{equation}
and
\begin{equation}
\label{3.6}
\partial_{x}^{(1)} n_{lx}(x,t) \equiv  n_{lx,eq}(x) \frac{\partial}{\partial x}\, \nu (x,t).
\end{equation}
Then, for instance, it is
\begin{eqnarray}
\label{3.7}
\frac{\partial^{2}}{\partial x^{2}} n_{lx}(x,t) &=& \nu(x,t) \frac{\partial^{2}}{\partial x^{2}}\, n_{lx,eq}(x) + 2 \epsilon \left[ 
\frac{\partial}{\partial x}\, n_{lx,eq}(x)  \right] \frac{\partial}{\partial x}\, \nu(x,t) \nonumber \\
&&+ \epsilon^{2} n_{lx,eq}(x) \frac{\partial^{2}}{\partial x^{2}} \nu(x,t) \nonumber \\
&=& \partial_{x}^{2(0)} n_{lx}(x,t)+ \epsilon \partial_{x}^{2(1)} n_{lx}(x,t) + \epsilon^{2} \partial_{x}^{2(2)} n_{lx}(x,t),
\end{eqnarray}
with the terms  $\partial_{x}^{2(i)} n_{lx}(x,t)$, $i=0,1,2$,  identified by comparison of both sides of the last equality. A general alternative Chapman-Enskog expansion about non-equilibrium states was formulated by Lutsko several years ago \cite{Lu06}. Although his general method can be easily applied to the problem addressed in this paper, it is more convenient to use an expansion based on the decomposition introduced by means of Eq. (\ref{3.4}), since it leads to a much simpler identification of the the zeroth order solution. A comparison of the expansion formulated by Lutsko and the one used in this paper is provided in Appendix A of ref. \cite{BGyM20}.

For a normal distribution, the above decomposition of the spatial derivatives of the density field, implies an expansion of the distribution function of the form
\begin{eqnarray}
\label{3.8}
f_{lx} \left[ {\bm v}|n_{lx}(x,t) \right] &=&  f_{lx}^{(0)} \left[ {\bm v}, \nu(x,t)|n_{lx,eq}(x) \right] + \epsilon  f_{lx}^{(1)}  \left[ {\bm v}, \nu(x,t), \frac{\partial\nu(x,t)}{\partial x}|n_{lx,eq}(x) \right] 
\nonumber \\
&&+ \epsilon^{2}   f_{lx}^{(2)} \left[ {\bm v}, \nu(x,t),\frac{\partial^{2} \nu(x,t)}{\partial x^{2}}, \left( \frac{\partial \nu(x,t)}{\partial x} \right)^{2}   |n_{lx,eq}(x) \right] + \ldots ,
\end{eqnarray}
where $f_{lx}^{(0)}$ is of zeroth order in $\partial \nu /\partial x$, $f_{lx}^{(1)}$ is linear in $\partial \nu / \partial x$, $f_{lx}^{(2)}$ contains terms that are  linear in $\partial^{2} \nu / \partial x^{2}
$ or in  $\left( \partial \nu / \partial x \right)^{2}$, and so on. At each order in the perturbation expansion, the distribution is a function of the exact density field through $\nu(x,t)$ and its spatial derivatives up to the corresponding order, as well as of all gradients of the equilibrium density profile, $n_{lx,eq}(x)$. As a consequence, there are two sources of space dependence in each term of the expansion in Eq.\ (\ref{3.8}): the scaling of the exact density field with the equilibrium one,  and the density of the equilibrium state itself. It is in the functional dependence of the normal solution to each order on the nonuniform equilibrium density field, where the difference with the usual Chapman-Enskog expansion shows up. Introduction of Eq.\ (\ref{3.8}) into the expression of the flux of labeled particles, Eq.\, (\ref{2.12}), directly leads to
\begin{equation}
\label{3.9}
J_{l}(x,t) = \sum_{j=0}^{\infty} \epsilon^{j} J_{l}^{(j)} (x,t),
\end{equation}
with
\begin{equation}
\label{3.10}
J_{l}^{(j)} (x,t) \equiv \int d{\bm v}\, v_{x} f_{lx}^{(j)} [{\bm v},\nu(x,t),\ldots |n_{lx,eq}(x)].
\end{equation}
 As already indicated, the physical motivation of the peculiar expansion being carried out here implies that, although $f_{lx}^{(0)}$ contains  gradients of all orders of $n_{lx,eq}(x)$, it must lead to a flux of labelled particles $J_{lx}^{(0)}=0$. Bellow it is verified that this is the case.
When the expansion in Eq.\ (\ref{3.9}) is considered on the right-hand-side of the conservation law, Eq.\, (\ref{2.11}), a similar expansion for  $\partial n_{lx}(x,t)/\partial t$ and hence for $\partial \nu(x,t)/\partial t$, is obtained. Due again to the normal property of the solution we are looking for, the latter allows for a multiescale  expansion of the time derivative of the $j$ order contribution to the distribution function,
\begin{equation}
\label{3.11}
\frac{\partial f_{lx}^{(j)}}{\partial t} = \partial_{t}^{(0)} f_{lx}^{(j)}+ \epsilon \partial_{t}^{(1)} f_{lx}^{(j)} + \epsilon^{2}
\partial_{t}^{(2)} f_{lx}^{(j)} + \ldots
\end{equation}
In the $\epsilon$-expansion of the distribution function, Eq. (\ref{3.8}), the lowest order contribution, $f_{lx}^{(0)} [{\bm v}|n_{lx}(x,t)]$, is defined such that it reproduces the actual density field of labeled particles along the $x$-axis, i.e.
\begin{equation}
\label{3.12}
\int d{\bm v}\ f_{lx}^{(0)} [{\bm v} ,\nu(x,t)|n_{lx, eq}(x)] = n_{lx}(x,t),
\end{equation}
while
\begin{equation}
\label{3.13}
\int d{\bm v}\ f_{lx}^{(j)} [{\bm v},\nu(x,t), \frac{\partial \nu (x,t)}{\partial t}, \ldots |n_{lx,eq}(x)] = 0,
\end{equation}
for $j \geq 1$. In this way, the expansion is consistent with  the definition of the density of labeled particles along the $x$-axis, Eq.\ (\ref{2.10}).
 
 The modified Chapman-Enskog expansion formulated above, will be used in the following to compute the flux of labeled particles, $J_{l} (x,t)$, defined in Eq. (\ref{2.12}),  to first order in the uniformity parameter $\epsilon$. Introduction of the obtained flux into the conservation law given in Eq. (\ref{2.11})  leads to the transport equation for labeled particles. Note that, formally, the gradient  operator in front of the particles flux in the conservation law could be also expanded by means of the decomposition introduced in Eq. (\ref{3.4}). If this were done, the exact law expressing the conservation of the number of labeled particles would be violated, being only verified up to the order in $\epsilon$ considered. Here, the relevance of the conservation law will  be taken into  account and the gradient operator in Eq. (\ref{2.11}) will be kept, without decomposing it.

Taking into account the several $\epsilon$-expansions indicated above, it follows that to zeroth order in $\epsilon$, the  kinetic equation (\ref{2.17}) reads
\begin{eqnarray}
\label{3.14}
&&\left( \partial_{t}^{(0)} + v_{x} \partial_{x}^{(0)} \right) f_{lx}^{(0)} [{\bm v}, \nu(x,t) |n_{lx,eq}(x)] -  v_{x} \frac{\partial \ln n_{lx,eq}(x)}{\partial x}\, f_{lx,0}(x,{\bm v},t) \nonumber \\ 
&&= g_{e}(n) \Lambda_{BL} [{\bm v}|f_{eq}] f_{lx}^{(0)} [{\bm v},\nu(x,t)| n_{lx,eq}(x)].
\end{eqnarray}
Moreover, the zeroth order balance equation for the density of labeled particles, Eq. (\ref{2.11}), is
\begin{equation}
\label{3.15}
\partial_{t}^{(0)} n_{lx}(x,t)+ \partial_{x}^{(0)} J_{l}^{(0)}(x,t)=0.
\end{equation}
The solution of Eq.\ (\ref{3.14}) accomplishing  the condition
\begin{equation}
\label{3.16}
\lim_{\nu(x,t) \rightarrow 1}  f_{lx}^{(0)} [{\bm v}, \nu(x,t) |n_{lx,eq}(x)] = f_{lx,eq}(x,{\bm v}),
\end{equation}
 is
\begin{equation}
\label{3.17}
f_{lx}^{(0)}[{\bm v}, \nu(x,t)|n_{lx,eq}(x)] = f_{lx,0}(x,{\bm v},t).
\end{equation}
To verify it, firstly note that  it is
\begin{equation}
\label{3.18}
J_{l}^{(0)} (x,t) = \int d{\bm v}\, v_{x} f_{lx,0} (x,{\bm v},t)=0
\end{equation}
and, hence,
\begin{equation}
\label{3.19}
\partial_{t}^{(0)} n_{lx}(x,t) =0.
\end{equation}
It follows that
\begin{equation}
\label{3.20}
\partial_{t}^{(0)} f_{lx,0}(x,{\bm v},t) = \left[ \partial_{t}^{(0)} n_{lx}(x,t)] \right] \varphi ({\bm v} )=0.
\end{equation}
In addition, it is 
\begin{eqnarray}
\label{3.21}
\partial_{x}^{(0)} f_{lx,0}(x,{\bm v},t) & = & \left[ \partial_{x}^{ (0)} n_{lx}(x,t) \right] \varphi({\bm v}) =\nu(x,t) 
\frac{\partial A(x)}{\partial x}  n_{l} \varphi({\bm v}) \nonumber \\
&= & \frac{ \partial \ln n_{lx,eq}(x)}{\partial x}\, f_{lx,0}(x,{\bm v},t)
\end{eqnarray}
and
\begin{equation}
\label{3.22}
\Lambda_{BL} [{\bm v}|f_{eq}] f_{lx}^{(0)} [{\bm v}, \nu(x,t)| n_{lx,eq}(x)] =0.
\end{equation}
The above proves that $f_{lx,0}(x,{\bm v},t)$ verifies Eq.\ (\ref{3.14}). Finally, the condition (\ref{3.12}) is trivially fulfilled. Therefore, the zeroth order distribution function of labeled particles, $f_{lx}^{(0)}$, in the modified Chapman-Enskog expansion we are using, is the local equilibrium one, i.e. it is obtained from the equilibrium distribution by replacing the equilibrium density profile, $n_{lx,eq}(x)$, by the actual non-equilibrium density field, $n_{lx}(x,t)$. The simplicity of this result is due to two reasons: the reference state is the equilibrium state, although it happens to be inhomogeneous, and the choice of the uniformity parameter $\epsilon$ as defined in Eq. (\ref{3.4}). 

\section{The macroscopic transport equation}
\label{s4}
By collecting the terms of first order in the uniformity parameter $\epsilon$ in Eq. (\ref{2.17}), the first order distribution function of labeled particles, $f_{lx}^{(1)}$, is found to obey the equation
\begin{eqnarray}
\label{4.1}
& & \left( \partial_{t}^{(0)} + v_{x} \partial_{x}^{(0)}  - g_{e}(n)  \Lambda_{BL}  \left[ {\bm v}  | f_{eq} \right] \right) f_{lx}^{(1)} \left[ {\bm v}, \nu(x,t), \frac{\partial \nu (x,t)}{\partial x} | n_{lx,eq} (x) \right]  \nonumber \\
& & = -\left( \partial_{t}^{(1)}  + v_{x} \partial_{x}^{(1)} \right) f_{lx}^{(0)} [{\bm v},\nu(x,t) |n_{lx, eq}(x)]
\end{eqnarray}
and the balance equation (\ref{2.11}) at the same order is
\begin{equation}
\label{4.2}
\partial_{t}^{(1)} n_{lx}(x,t) +\partial_{x}^{(0)} J_{l}^{(1)}(x,t) =0,
\end{equation}
where Eq.\ (\ref{3.18}) has been employed. Evaluation of the several terms in Eq. (\ref{4.1}), using the rules developed in the previous section, yields
\begin{eqnarray}
\label{4.3}
&& v_{x} \int dx^{\prime}\, \ \frac{\partial A(x^{\prime})}{\partial x^{\prime}}  n_{l} \frac{\delta f_{lx}^{(1)}}{\delta n_{lx,eq}(x^{\prime})} - \left[ \partial_{x}^{(0)} J_{l}^{(1)}(x,t) \right] \varphi ({\bm v}) - g_{e}(n) \Lambda_{BL} \left[ {\bm v}|f_{eq} \right] f_{lx}^{(1)} \nonumber \\
&&  =- v_{x} n_{lx,eq}(x) \varphi ({\bm v})  \frac{\partial \nu (x,t)}{\partial x}.
\end{eqnarray}
The solution of this equation must have the form
\begin{equation}
\label{4.4}
f_{lx}^{(1)} \left[ {\bm v}, \nu(x,t), \frac{\partial \nu (x,t)}{\partial x} | n_{lx,eq} (x) \right]= B \left[ {\bm v} , \nu(x,t) |n_{lx,eq}(x)\right]  n_{lx,eq}(x) \frac{\partial \nu (x,t)}{\partial x}.
\end{equation}
Here, $B$ is a function of ${\bm v}$ and $\nu(x,t)$, and a functional of $n_{lx,eq}(x)$, although it will not be always explicitly indicated in the notation.  Then, the flux of labeled particles along the $x$ direction to first order in the uniformity parameter has the form
\begin{equation}
\label{4.5}
J_{l}^{(1)}(x,t)= - D(x,t) \frac{\partial}{\partial x}\, \nu(x,t),
\end{equation}
with the self-diffusion coefficient $D(x,t)$ given by
\begin{equation}
\label{4.6}
D(x,t)\equiv  -n_{lx,eq}(x) \int d{\bm v}\, v_{x} B \left[ {\bm v} , \nu(x,t) |n_{lx,eq}(x)\right] .
\end{equation}
Substitution of Eq.\ (\ref{4.4}) into Eq. (\ref{4.3}), taking into account the arbitrariness of $\partial \nu /\partial x$, gives 
\begin{eqnarray}
\label{4.7}
&&  \int dx^{\prime}\,  \frac{\partial A(x^{\prime})}{\partial x^{\prime}}  n_{l} \frac{\delta}{\delta n_{lx,eq}(x^{\prime})}  \left[ B({\bm v}) n_{lx,eq}(x) v_{x} + D(x,t) \varphi({\bm v}) \right]  \nonumber  \\ 
&& - g_{e}(n)  \Lambda_{BL}  \left[ {\bm v}  | f_{eq} \right] B({\bm v}) n_{lx.eq}(x) = - v_{x} n_{lx,eq}(x) \varphi({\bm v}).
\end{eqnarray} 
Integration of both sides of this equation over ${\bm v}$ leads to a trivial identity that does not provide any information about $B({\bm v})$ or the self-diffusion coefficient $D(x,t)$. Symmetry considerations 
outlined in Appendix \ref{ap2} show that a lowest order Sonine approximation  \cite{RydL77} for  $B({\bm v})$ is given by
\begin{equation}
\label{4.8}
B({\bm v})= \varphi ({\bm v}) \left[ a_{01} \left( \frac{1}{2} - \frac{mv_{y}^{2}}{2k_{B}T} \right) + a_{10} \left( \frac{1}{2} - \frac{mv_{x}^{2}}{2k_{B}T} \right) + b_{00} \left( \frac{m}{2k_{B}T} \right)^{1/2} v_{x} \right],
\end{equation}
where $a_{01}$, $a_{10}$, and $b_{00}$ are velocity-independent quantities to be determined in the following. They can be functions of $\nu (x,t)$ and functionals of $n_{lx,eq} (x)$. In this approximation,
Eq.\, (\ref{4.6}) becomes
\begin{equation}
\label{4.9}
D(x,t) \approx D(x)= -n_{lx,eq}(x) \left( \frac{k_{B}T}{m} \right)^{1/2} b_{00}.
\end{equation}
 Substitution of Eq. (\ref{4.8}) in Eq.\, (\ref{4.7}), multiplication of both sides of the equation by $v_{x}$, and integration over ${\bm v}$, after some algebra leads to
 \begin{equation}
 \label{4.10}
 n_{l}  \int dx^{\prime}\,  \frac{\partial A(x^{\prime})}{\partial x^{\prime}} \frac{\delta}{\delta n_{lx,eq}(x^{\prime})} \left[ a_{10} n_{lx,eq} (x) \right] - \sqrt{2 \pi} n g_{e}(n) b_{00} n_{lx,eq}(x) = n_{lx,eq}(x).
 \end{equation}
To close this equation and determine $b_{00}$, that is needed to get the expression for the self-diffusion coefficient $D$, additional independent equations for $a_{01}$, $a_{10}$, and $b_{00}$ are required. Multiplication of Eq. (\ref{4.7}) by $v_{x}^{2} $, after substituting the expression for $B({\bm v})$ given in Eq. (\ref{4.8}), and later integration over the velocity gives
\begin{equation}
\label{4.11}
n_{l}  \int dx^{\prime}\, \frac{\partial A(x^{\prime})}{\partial x^{\prime}} \frac{\delta}{\delta n_{lx,eq}(x^{\prime})} \left[ b_{00} n_{lx,eq} (x) \right] - \frac{1}{2} \sqrt{\frac{\pi}{2}} \left( 5a_{10} -a_{01} \right) g_{e} (n) n \sigma n_{lx,eq}(x)=0.
\end{equation}
Finally, multiplication of Eq. (\ref{4.7}) by $v_{y}^{2}$, use of Eq. (\ref{4.8}), and integration over ${\bm v}$ yields
\begin{equation}
\label{4.12}
5a_{01}= a_{10}.
\end{equation}
In this way, the set of closed equations (\ref{4.10}), (\ref{4.11}), and (\ref{4.12})  for the quantities $a_{01}$, $a_{10}$, and $b_{00}$ has been obtained. They are functional equations rather difficult to solve in an exact way, so an iterative approximation will be employed. First, Eq. (\ref{4.10}) is considered with $a_{10}  \approx a_{10}^{(0)}=0$. Because of Eq. (\ref{4.12}), this implies $a_{01} \approx a_{01}^{(0)} =0$. The solution is
\begin{equation}
\label{4.13}
b_{00} \approx b_{00}^{(0)} =  - \frac{1}{\sqrt{2\pi} n \sigma g_{e}(n)}.
\end{equation}
Next, $b_{00}$ is used into Eqs. (\ref{4.11}) and (\ref{4.12}) to get first order corrections, $a_{01}^{(1)}$ and $a_{10}^{(1)}$ for $a_{01}$ and $a_{10}$. Finally, $a_{10}^{(1)} $ is employed in Eq. (\ref{4.10}) to obtain a first order expression, $b_{00}^{(1)}$, for $b_{00}$. When this approximation scheme is carried out, the result reads
\begin{equation}
\label{4.14}
a_{10}^{(1)} =- \frac{5}{12 \pi}\, \frac{1}{ \left[ g_{e}(n)  n \sigma \right]^{2}} \frac{\partial}{\partial x} \ln n_{lx,eq}(x),
\end{equation}
\begin{equation}
\label{4.15}
b_{00}^{(1)} =  b_{00}^{(0)} + \frac{5 b_{00}^{(0)3}}{6 n_{lx,eq}(x)} \frac{\partial^{2}}{\partial x^{2}}\, n_{lx,eq}(x).
\end{equation}
The expression for $a_{01}^{(1)}$ follows by means of  Eq.\ (\ref{4.12}). Use of Eq.\ (\ref{4.15}) into Eq. (\ref{4.9}) provides the result for the self-diffusion coefficient,
\begin{equation}
\label{4.16}
D(x) \approx D^{(1)}(x) =  D^{(0)}(x) + D^{\prime} (x),
\end{equation}
where
\begin{equation}
\label{4.17}
D^{(0)} (x) = n_{lx,eq}(x) D_{0}, \quad D_{0} =  \ell_{0} \left( \frac{k_{B}T}{m} \right)^{1/2}\, ,
\end{equation} 
and 
\begin{equation}
\label{4.18}
D^{\prime} (x)= \frac{5}{3} \ell_{0}^{3} \left( \frac{k_{B} T}{m} \right) ^{1/2} \frac{\partial^{2}}{\partial x^{2}}\, n_{lx,eq}(x) = \frac{5}{3} D_{0} \ell_{0 }^{2} \frac{\partial^{2}}{\partial x^{2}} n_{lx,eq}(x).
\end{equation}
In the above expressions,  $\ell_{0}$ is a measurement of the mean-free-path of the particles in the gas,
\begin{equation}
\label{4.19}
\ell_{0} \equiv  \frac{1}{2 \sqrt{\pi} n \sigma g_{e}(n)}\, .
\end{equation}
The coefficient $D_{0}$ agrees with the bulk equilibrium self-diffusion coefficient for a two-dimensional system of hard disks obtained from the Enskog equation \cite{Gas71}. 
In summary, it has ben found that the flux of labeled particles to first order in its effective  density gradient  is given by (see Eq.\ (\ref{4.5}))
\begin{equation}
\label{4.20}
J_{l}^{(1)}(x,t) \approx -  D^{(1)} (x) \frac{\partial}{\partial x}\, \nu(x,t)
\end{equation}
and, consequently, the transport equation for labeled particles to the same order is
\begin{equation}
\label{4.21}
\frac{ \partial n_{lx}(x,t)}{\partial t} = \frac{\partial}{\partial x} \left[ D^{(1)}(x) \frac{\partial}{\partial x}\, \nu (x,t) \right].
\end{equation}
This has the form of a modified FJ equation (\ref{1.3}), with an effective longitudinal diffusion coefficient 
\begin{equation}\
\label{4.22}
D^{(1)} (x) = D_{0} \left[ n_{lx,eq}(x)+ \frac{5}{3} \ell_{0 }^{2} \frac{\partial^{2}}{\partial x^{2}} n_{lx,eq}(x) \right].
\end{equation}

It is interesting to analyze the form of the first order correction to the local equilibrium distribution function in the modified Champan-Enskog expansion that has been developed here. From Eqs.\, (\ref{4.4}), (\ref{4.8}), (\ref{4.12}), (\ref{4.14}), and (\ref{4.15}) it is found
\begin{equation}
\label{4.23}
f_{lx}^{(1)} (x,{\bm v},t) =- \varphi ({\bm v}) \left\{ \ell_{0}^{2} \left[ 1- \frac{m}{6k_{B}T} \left ( v_{y}^{2} +5v_{x}^{2} \right) \right] \frac{\partial n_{lx,eq}(x)}{\partial x} + D^{(1)} (x) \frac{mv_{x}}{k_{B}T} \right\} \frac{\partial \nu (x,t)}{\partial x}\, .
\end{equation}
From this expression and Eq.\ (\ref{3.17}), the second moments of the velocity components can be  directly computed to first order in $\partial \nu / \partial x$,
\begin{eqnarray}
\label{4.24}
\langle v_{x}^{2} \rangle &\equiv& \frac{1}{n_{lx}(x,t)}  \int d{\bm v}\, v_{x}^{2}\left[f_{lx}^{(0)}(x,{\bm v},t)+ f_{lx}^{(1)} (x,{\bm v},t) \right] \nonumber \\
&= & \frac{k_{B}T}{m} \left[ 1 + \frac{5 \ell_{0}^{2}}{3 n_{lx}(x,t)} \frac{\partial n_{lx,eq}(x)}{\partial x} \frac{\partial \nu (x,t)}{\partial x} \right],
\end{eqnarray}
\begin{eqnarray}
\label{4.25}
\langle v_{y}^{2} \rangle &\equiv& \frac{1}{n_{lx}(x,t)}  \int d{\bm v}\, v_{y}^{2}\left[f_{lx}^{(0)}(x,{\bm v},t)+ f_{lx}^{(1)} (x,{\bm v},t) \right] \nonumber \\
&= & \frac{k_{B}T}{m} \left[ 1 + \frac{ \ell_{0}^{2}}{3 n_{lx}(x,t)} \frac{\partial n_{lx,eq}(x)}{\partial x} \frac{\partial \nu (x,t)}{\partial x} \right].
\end{eqnarray}
If local temperature parameters, $T_{l,x}$ and $T_{l,y}$, associated to each of the degrees of freedom are defined for the labeled particles by 
\begin{equation}
\label{4.26}
T_{l,x}(x,t) \equiv  \frac{m \langle v_{x}^{2} \rangle}{k_{B}} , \quad T_{l,y}(x,t) \equiv  \frac{m \langle v_{y}^{2} \rangle}{k_{B}}\, ,
\end{equation} 
the above results show that, in general  $T_{l,x} \neq T_{l,y}$. This effect is a consequence of both, a varying cross section of the channel and the presence of a gradient of labeled particles relative to its equilibrium value. Actually, if a global local temperature of the labeled particles is defined as
\begin{equation}
\label{4.27}
T_{l}(x,t) = \frac{T_{l,x}(x,t) + T_{l,y}(x,t)}{2},
\end{equation}
it is
\begin{equation}
\label{4.28}
T_{l}(x,t) = T \left[ 1+  \frac{ \ell_{0}^{2}}{ n_{lx}(x,t)} \frac{\partial n_{lx,eq}(x)}{\partial x} \frac{\partial \nu (x,t)}{\partial x} \right],
\end{equation}
and whether  the temperature of the labeled particles is  larger or smaller than the temperature of the bath, depends on the shape of both, the equilibrium and non-equilibrium densities of labeled particles.

\section{Conclusions}
\label{s5}
In this paper, the problem of one-dimensional transport along the longitudinal direction in a channel with a varying transversal width has been addressed, starting from a description of the system at the level of kinetic theory. The simplest case of self-diffusion in a two-dimensional system has been considered, but the method can also be  applied to more complex transport situations and also to three-dimensional systems. 

As a consequence of the dependence of the cross section on the longitudinal coordinate, the equilibrium number density of particles along that direction is inhomogeneous. That implies that not all the density gradients along the longitudinal direction can be expected to generate a flux of particles. This leads to consider a modified Chapman-Enskog expansion method when generating hydrodynamics from the starting kinetic equation. The uniformity parameter is not associated to the gradient of the actual density field, but to the gradient of the density field scaled  by its equilibrium value.  

The usual macroscopic equation to describe one-dimensional diffusion in a channel of varying cross section is known as the Fick-Jacobs equation  and it was originally derived by heuristic methods \cite{Ja67,Fi55}. Later on, it was obtained by coordinate reduction from a macroscopic equation in two (or three) dimensions. More concretely, Zwanzig  \cite{Zw92} used the Smoluchowski equation for diffusion in an external conservative field  generating the hard walls confining the system, while Kalinay and Perkus  \cite{KyP06} employed an anisotropic  diffusion equation supplemented with the appropriated boundary conditions.  

In refs. \cite{Zw92} and \cite{KyP06}, corrections to the Ficks-Jacobs equation are derived for the case of mirror or cylindrical symmetry. They follows from considering density profiles in the transversal direction and can be expressed in the one-dimensional equation by means of a position dependent diffusion coefficient, that involves the local curvature of the channel.
A similar description has been proposed in the context of a mesoscopic non-equilibrium thermodynamics approach to the problem \cite{RyR01}. Although the result reported in this paper can also be expressed by means of a position dependent diffusion coefficient, the nature of the present study and its results are  rather different from the previous ones, where a fundamental role is played by the fact that the transversal dynamics also obeys the macroscopic diffusion equation, something that has not been 
assumed in the present study. Actually, also the symmetry of the system is crucial for deriving the results reported in those works. 
The correction to the Jacobs-Fick law derived here has a mathematical and physical origin that is different from those mentioned above. It follows from the fact that only gradients of the deviations from the equilibrium density generate macroscopic fluxes of particles in the system. Moreover, the analysis carried out also predicts the  anisotropy of the temperature parameters associated to the vertical and horizontal motions.

To put the present work in a proper context, it must be emphasized that the obtained transport equation is not an exact consequence of the starting model kinetic equation formulated in Eq.\, (\ref{2.17}) and the modified Chapman-Enskog expansion. Two main approximations have been introduced. First, a Sonine expansion of the distribution function has been considered, and only first order terms have been kept. This is expected to provide an accurate description, as long as the distribution function of the labeled particles be close to a Maxwellian. The second approximation made was to solve the system of functional equations
obeyed by the coefficients defining the first Sonine order, by means of an iterative procedure. Although this point has not been investigated in detail, it seems plausible that this procedure imply neglecting high order gradients of the longitudinal density equilibrium distribution. The expectation is that corrections associated to those derivatives be much smaller than the one that has been included.

It seems relevant to remark that the derived results do not apply in the limit of a strongly confined gas, i.e. when the distance between the two hard boundaries is slightly larger than the diameter of the particles. In this case, it is necessary to modify the kinetic equation (\ref{2.3}) by incorporating the limitations on the posible values of the collision angle between two particles, as a consequence of the imposed confinement \cite{BMyG16, BGyM17,MGyB18}.  Nevertheless, the present analysis can be useful to investigate diffusion in nanosystems, in which the width of the system is much larger than the size of the particles but smaller than the characteristic length needed for the validity of hydrodynamics \cite{SHyR08,ByC10}.

The validation of the modified Ficks-Jacobs equation obtained here can not be done by comparison with the results obtained from the exact solution of the the macroscopic two-dimensional diffusion equation, known for some specific geometries, as it is done in refs. \cite{Zw92,KyP06,RyR01}. The point raised in the present study is that the two-dimensional macroscopic equation may not lead, by means of a coordinate reduction procedure, to a consistent macroscopic equation for the time evolution of the density along the longitudinal direction. The reason is the change in the appropriate definition of the uniformity parameter, as discussed along this paper and, especially in Sec. \ref{s3}.  Moreover, it is clear that starting from  a diffusion equation nothing can be deduced about the possible anisotropy of the temperature parameters, an effect that has been identified here. Of course, it seems quite interesting to check  the predictions presented in this paper by comparing with molecular dynamics simulation results, something we are carrying out currently. 

\section*{Funding}

This research was supported by grant ProyExcel-00505 funded by Junta de Andaluc\'{\i}a and GrantPID2021-126348N funded by MCIN/AEI/10.13039/501100011033 and "ERDF A way of making Europe".

\appendix

\section{The boundary conditions at the hard walls and the derivation of Eq. (\ref{2.11})}
\label{ap1}
Consider the curve $y=h(x)$. The system is below it and the unit vector normal to the curve, $\widehat{\bm e}_{n}$ is defined as directed out of the system. Then, it is
\begin{equation}
\label{ap1.1}
\widehat{\bm e}_{n}= \frac{1}{\left[1+h^{\prime}(x) \right]^{1/2}}\left[ -h^{\prime}(x) {\bm i}+{\bm j} \right],
\end{equation}
where ${\bm i}$ and ${\bm j}$ are the unit vectors along the $x$ and $y$ axis, respectively and $h^{\prime}(x) \equiv \partial h(x) /\partial x$. The boundary condition corresponding to an elastic hard wall at this curve is \cite{DyvB77}
\begin{equation}
\label{ap1.2}
\delta \left[ y-f(x) \right] \Theta \left( {\bm v} \cdot \widehat{\bm e}_{n} \right) | {\bm v} \cdot \widehat{\bm e}_{n}| f_{l}({\bm r},{\bm v},t) = \delta \left[y-f(x) \right] \Theta \left( {\bm v} \cdot \widehat{\bm e}_{n} \right) | {\bm v} \cdot \widehat{\bm e}_{n}| f_{l}({\bm r},{\bm v}^{*},t).
\end{equation}
In this expression, ${\bm v}^{*}$ is the velocity of a particle after colliding with the wall, being ${\bm v}$ its velocity prior the collision,
\begin{equation}
\label{ap1.3}
{\bm v}^{*}= {\bm v}- 2 {\bm v} \cdot \widehat{\bm e}_{n}  \widehat{\bm e}_{n} .
\end{equation}
To study the boundary located at $y=g(x)$, a similar analysis is carried out. The main difference is that now the system is below the curve. The result is
\begin{equation}
\label{ap1.4}
\delta \left[ y-g(x) \right] \Theta \left( - {\bm v} \cdot \widehat{\bm e}_{n^{\prime}} \right) | {\bm v} \cdot \widehat{\bm e}_{n^{\prime}}| f_{l}({\bm r},{\bm v},t) = \delta \left[y-g(x) \right] \Theta \left( -{\bm v} \cdot \widehat{\bm e}_{n^{\prime}} \right) | {\bm v} \cdot \widehat{\bm e}_{n^{\prime}}| f_{l}({\bm r},{\bm v}^{**},t),
\end{equation}
with
\begin{equation}
\label{ap1.5}
{\bm v}^{**}= {\bm v}+ 2 {\bm v} \cdot \widehat{\bm e}_{n^{\prime}} \widehat{\bm e}_{n^{\prime}}
\end{equation}
and
\begin{equation}
\label{ap1.6}
\widehat{\bm e}_{n^{\prime}}= \frac{1}{\left[1+g^{\prime}(x)^{2} \right]^{1/2}}\left[ -g^{\prime}(x) {\bm i}+{\bm j} \right],
\end{equation}
$g^{\prime}(x) \equiv \partial h(x) /\partial x $. Consider now the integration over the velocity of Eq. (\ref{2.9})  that yields
\begin{eqnarray}
\label{ap1.7}
\frac{\partial}{\partial t}\, n_{lx}(x,t) + \frac{\partial}{\partial x}\, J_{l}(x,t) &- & \int d{\bm v} \left[ v_{x} h^{\prime} (x)- v_{y} \right] f_{l}(x,h(x),{\bm v},t)  \nonumber \\
&+ &\int d{\bm v} \left[ v_{x} g^{\prime}(x)- v_{y} \right] f_{l}(x,g(x),{\bm v},t)=0,
\end{eqnarray}
where the property given in Eq.\ (\ref{2.6}) has been employed. From Eq. (\ref{ap1.1}) it follows that
\begin{equation}
\label{ap1.8}
v_{x} h^{\prime} (x)- v_{y} = -\widehat{\bm e}_{n} \cdot {\bm v} \left[1+h^{\prime}(x)^{2} \right]^{1/2}.
\end{equation}
Then, taking into account the boundary condition  in Eq. (\ref{ap1.2}), it is found
\begin{eqnarray}
\label{ap1.9}
&& \int d{\bm v}\ \left[v_{x} h^{\prime} (x)- v_{y} \right] f_{l}(x,h(x),{\bm v},t) =-  \left[1+h^{\prime}(x)^{2} \right]^{1/2} \int d{\bm v}\, \widehat{\bm e}_{n} \cdot {\bm v} f_{l}(x,h(x),{\bm v},t) \nonumber \\
&& = - \left[1+h^{\prime}(x)^{2} \right]^{1/2}  \int_{-\infty}^{+\infty} dy \int d{\bm v}\,  \delta \left( y-h(x) \right) \widehat{\bm e}_{n} \cdot {\bm v} f_{l}(x,y,{\bm v},t) \nonumber \\
&& = -  \left[1+h^{\prime}(x)^{2} \right]^{1/2}  \int_{-\infty}^{+\infty} dy \int d{\bm v}\,  \delta \left( y-h(x) \right) \left[ \Theta (\widehat{\bm e}_{n} \cdot {\bm v} ) +  \Theta (-\widehat{\bm e}_{n} \cdot {\bm v} ) \right] \nonumber \\
&& \times \widehat{\bm e}_{n} \cdot {\bm v} f_{l}(x,y,{\bm v},t) \nonumber \\
&& = - \left[1+h^{\prime}(x)^{2} \right]^{1/2}  \int_{-\infty}^{+\infty} dy \int d{\bm v}\,  \delta \left( y-h(x) \right)  \Theta (\widehat{\bm e}_{n} \cdot {\bm v}) \widehat{\bm e}_{n} \cdot {\bm v} f_{l}(x,y,{\bm v}^{*},t) \nonumber \\
&& - \left[1+h^{\prime}(x)^{2} \right]^{1/2}  \int_{-\infty}^{+\infty} dy \int d{\bm v}\,  \delta \left( y-h(x) \right) \Theta (-\widehat{\bm e}_{n} \cdot {\bm v}) \widehat{\bm e}_{n} \cdot {\bm v} f_{l}(x,y,{\bm v},t).
\end{eqnarray}
But, it is
\begin{equation}
\label{ap1.10}
 \int d{\bm v}\,  \Theta (\widehat{\bm e}_{n} \cdot {\bm v}) \widehat{\bm e}_{n} \cdot {\bm v} f_{l}(x,y,{\bm v}^{*},t) =  - \int d{\bm v}\,  \Theta (-\widehat{\bm e}_{n} \cdot {\bm v}) (-\widehat{\bm e}_{n} \cdot {\bm v}) f_{l}(x,y,{\bm v},t),
\end{equation}
and, consequently, the third term on the left hand side of Eq. (\ref{ap1.7}) vanishes. A similar calculation holds for the last term on the left hand side of the same equation, so that the equation  reduces to Eq.\ (\ref{2.11}).

\section{First Sonine approximation for $B({\bm v})$, solution of Eq. (\ref{4.7})}
\label{ap2}
Dimensional analysis requires that 
\begin{equation}
\label{ap2.1}
B({\bm v}) = \xi ({\bm c}),
\end{equation}
with ${\bm c} \equiv \left( k_{B}T/m \right)^{1/2}$. Moreover, the structure of Eq. (\ref{4.7}) implies that $\xi$ must be an even function of $c_{y}$. Due to normalization, it must be
\begin{equation}
\label{ap2.2}
\int d{\bm c}\, \xi ({\bm c})=0.
\end{equation}
Introduce the decomposition
\begin{equation}
\label{ap2.3}
\xi ({\bm c}) = \xi_{even} ({\bm c}) +\xi_{odd} ({\bm c}),
\end{equation}
where
\begin{equation}
\label{ap2.4}
\xi_{even} ({\bm c}) = \frac{ \xi({\bm c}) + \xi (- {\bm c})}{2} = \frac{ \xi(c_{x},c_{y}) + \xi (- c_{x}, c_{y})}{2}
\end{equation}
is an even function of $c_{x}$ and  $c_{y}$, and
\begin{equation}
\label{ap2.5}
\xi_{odd} ({\bm c}) = \frac{ \xi({\bm c}) - \xi (- {\bm c})}{2} =  \frac{ \xi(c_{x},c_{y}) - \xi (- c_{x}, c_{y})}{2}
\end{equation}
is and odd function of $c_{x}$ and an even function of $c_{y}$ . The function $\xi_{even} ({\bm c}) $ can be expanded in Sonine polynomials as \cite{RydL77},
\begin{eqnarray}
\label{ap2.6}
\xi_{even} ({\bm c})  =  Z(c_{x}^{2},c_{y}^{2}) & =  &\pi^{-1/2} \sum_{j=0}^{\infty} a_{j}(c_{x}^{2}) S_{-1/2}^{(j)} (c_{y}^{2}) e^{-c_{y}^{2}} \nonumber \\
&=& \pi^{-1} \sum_{i=0}^{\infty}  \sum_{j=0}^{\infty} a_{ij} S_{-1/2}^{(i)} (c_{x}^{2})  S_{-1/2}^{(j)} (c_{y}^{2}) e^{-c^{2}}.
\end{eqnarray}
Similarly,
\begin{equation}
\label{ap2.7}
\xi_{odd} ({\bm c})= c_{x}  Y(c_{x}^{2},c_{y}^{2}) = \pi^{-1}\sum_{i=0}^{\infty}  \sum_{j=0}^{\infty} b_{ij}  c_{x} S_{-1/2}^{(i)} (c_{x}^{2})  S_{-1/2}^{(j)} (c_{y}^{2}) e^{-c^{2}}.
\end{equation}
Note that the coefficients $a_{ij}$ and $b_{ij}$ are velocity independent. The Sonine polynomial $S_{-1/2}^{(i)}(z)$ are defined as \cite{RydL77}
\begin{equation}
\label{ap2.8}
S_{-1/2}(z) \equiv \sum_{k=0}^{i} \frac{ (-1)^{k} \Gamma (i+1/2)}{\Gamma (k+1/2) (i-k)! k!}\, z^{K}
\end{equation}
and satisfy the orthogonality condition
\begin{equation}
\label{ap2.9}
\int_{0}^{\infty} dz\, z^{-1/2} e^{-z} S_{-1/2}^{(i)} (z) S_{-1/2}^{(j)} (z) = \frac{\Gamma (i+1/2)}{i!}\, \delta_{ij}.
\end{equation}
In the above expressions $\Gamma (n)$ denota de Gamma Euler function. In particular, it is
\begin{equation}
\label{ap2.10}
S_{-1/2}^{(0)} = 1, \quad S_{-1/2}^{(1)} (z) = \frac{1}{2} -z .
\end{equation}
It is easily seen that the condition expressed in Eq. (\ref{ap2.2}) is equivalent to $a_{00}=0$. Taking this into account, the first Sonine approximation for $\xi({\bm c})$ is defined as
\begin{equation}
\label{ap2.11}
\xi({\bm c} )\approx \pi^{-1} e^{-c^{2}} \left[ a_{01} \left ( \frac{1}{2} -c_{y}^{2} \right) +a_{10} \left( \frac{1}{2} - c_{x}^{2} \right) + b_{00} c_{x} \right].
\end{equation}
Returning to the original velocity scale ${\bm v}$, this is recognized as Eq.\, (\ref{4.8}).  In order to have some information about the physical meaning of the coefficients $a_{01}$, $a_{10}$, and $b_{00}$, it is interesting to consider the first few moments of $\xi ({\bm c})$. It is 
\begin{equation}
\label{ap2.12}
\int d{\bm c}\, c_{y} \xi({\bm c})=0,
\end{equation}
\begin{equation}
\label{ap2.13}
\int d{\bm c}\, c_{x} \xi ({\bm c})= \frac{b_{00}}{2},
\end{equation}
\begin{equation}
\label{ap2.14}
\int d{\bm c}\, c_{y}^{2} \xi ({\bm c}) = - \frac{a_{01}}{2},
\end{equation}
\begin{equation}
\label{ap2.15}
\int d{\bm c}\, c_{x}^{2} \xi ({\bm c}) = -\frac{a_{10}}{2}.
\end{equation}

\end{document}